\documentclass[12pt]{article}

\usepackage{cite}
\usepackage{subfigure}
\usepackage{multirow}
\usepackage{helvet}
\usepackage{amsmath}
\usepackage{amssymb}
\usepackage{setspace}
\usepackage{setspace}
\usepackage[dvips]{graphicx}
\usepackage{epsfig}
\usepackage{empheq}

\begin{document}

\titlepage                                                    
\begin{center}                                                    
{\Large \bf Planar radiation zeros in five--parton QCD amplitudes}\\

 
 \vspace*{1cm}
                                                    
 L.A. Harland--Lang\\                                                 
                                                    
 \vspace*{0.5cm}                                                    
Department of Physics and Astronomy, University College London, WC1E 6BT, UK                                          
                                                     
 \vspace*{1cm}                                                    
 \begin{abstract}                                                    
 \noindent
  We demonstrate the existence of `planar', or `type--II', radiation zeros in 5--parton QCD scattering processes. That is, the Born amplitudes are shown to completely vanish for particular kinematic configurations, when all the particle 3--momenta lie in a plane. This result is shown to follow particularly simply from the known `BCJ' relations between the colour--ordered tree amplitudes, and the MHV formalism is used to express the additional kinematic constraint as a relatively simple expression in terms of only rapidity differences between the final--state partons. 
  In addition, we find that zeros exist for non--planar configurations of the final--state partons, but for which the normal `type--I' conditions on the particle four--momenta do not generally apply. We present numerical results and comment on the possibility of observing planar radiation zeros in hadronic collisions, via central exclusive three--jet production.

 \end{abstract}                                                        
 \vspace*{0.5cm}                                                    
\end{center}

\section{Introduction}

Processes involving the radiation of one or more massless gauge bosons are known to exhibit an interesting feature: the Born--level scattering amplitudes can completely vanish for particular configurations of the final--state particles, independent of their polarizations. This phenomenon, known as a `radiation zero', was discovered in~\cite{Brown:1979ux,Mikaelian:1979nr} in the electroweak process $u \overline{d} \to W^+\gamma$. Following this initial observation, a deeper theoretical understanding was developed;
it was found that  these zeros may occur due to the complete destructive interference of classical radiation patterns in any gauge--theory amplitude with massless gauge boson emission, and a general theorem for their existence was derived in~\cite{Brown:1982xx,Brodsky:1982sh,Samuel:1983eg}.
  This has lead to a range of phenomenological work (see~\cite{Han:1995ef,Brown:1995pya,Baur:1999ym} for reviews and references), and radiation zeros have been observed experimentally by the D0 collaboration at the Tevatron~\cite{Abazov:2008ad}, and more recently the CMS collaboration at the LHC~\cite{Chatrchyan:2013fya}, in the same $W\gamma$ channel that lead to their discovery some 30 years previously.

The general theorem derived in~\cite{Brown:1982xx,Brodsky:1982sh,Samuel:1983eg} can be understood most simply in the case of soft photon emission. Here, the radiative matrix element is given by
\begin{equation}\label{msoft}
\mathcal{M^\gamma}\approx e\, \epsilon \cdot J \mathcal{M}_0\;,
\end{equation}
where $\mathcal{M}_0$ is the matrix element without photon--emission and $\epsilon$/$k$ is the photon polarisation vector/4--momentum. The current $J$ is given by
\begin{equation}\label{jsoft}
J^\mu=\sum_i e_i\eta_i\frac{p_i^\mu}{p_i \cdot k}\;,
\end{equation}
where $e_i$ is the electric charge of the  $i$th particle and $\eta_i = +1$, $-1$ for incoming, outgoing particles. It then readily follows from 4--momentum conservation that the amplitude will vanish when all charge to light--cone--energy fractions are equal, that is
\begin{equation}\label{type1}
\frac{Q_i}{p_i \cdot k}=\kappa\;, 
\end{equation}
where the notation has been adjusted slightly for generality, and $\kappa$ is some constant, independent of $i$. Here $p_i$/$Q_i$ are the 4--momenta/charges of the remaining particles in the scattering process; for photon emission $Q_i$ are simply the particle electric charges as in (\ref{msoft}), while in the QCD case these correspond to suitable colour factors due to the gluon emission, see~\cite{Brodsky:1982sh}. While the soft limit is considered above, it can in fact be shown that provided (\ref{type1}) holds, then a zero will occur for arbitrary energies of the emitted gauge boson~\cite{Brown:1982xx}. 

However, in~\cite{Heyssler:1997fy} a new `type--II' zero was observed in the $e q \to eq \gamma$ process, the presence of which was not determined by the requirement (\ref{type1}) for the original `type--I' zeros. Indeed, in~\cite{Heyssler:1997ng}, similar zeros were shown to occur in the $e^+ e^- \to q\overline{q}\gamma$ process; in this case (\ref{type1}) can in fact never be satisfied, due to the differing signs of the particle electric charges. A straightforward necessary condition for such zeros was found: the scattering should be \emph{planar}, that is the 3--momenta of all scattered particles, including the emitted gauge boson, must lie in a plane. This requirement can be understood quite simply for the case of soft photon emission, where by choosing one photon polarisation vector to be orthogonal to the reaction plane it readily follows from (\ref{msoft}) and (\ref{jsoft}) that the scattering amplitude vanishes automatically. The remaining requirement that the amplitude vanishes for the polarization vector lying in the plane then defines an additional constraint on the final--state momenta which must be satisfied for a radiation zero to occur for arbitrary polarisations.
Other examples of these planar zeros can be found in~\cite{Stirling:1999sj,Rodriguez:2003tt}, however, these remain relatively unstudied, and the general principles for their occurrence has yet to be truly understood, in particular away from the soft limit. Moreover, until now the only general examples of these are found to exist in relatively simple QED processes, with no non--abelian vertices contributing; while~\cite{Stirling:1999sj} demonstrates the existence of planar zeros in the $e^+e^- \to W^+ W^- \gamma$ process, these are only found to occur in the soft photon limit.

 In this paper we demonstrate, for the first time, the existence of planar radiation zeros in QCD processes for general particle momenta\footnote{As discussed further in Section~\ref{conclusions}, radiation zeros have been observed in the $gg\to\pi\pi$~\cite{HarlandLang:2011qd} and  $gg\to J/\psi J/\psi$~\cite{Harland-Lang:2014efa}  processes, i.e. in the 6--parton $gg\to q\overline{q} q\overline{q}$ amplitudes where the collinear $q\overline{q}$ pairs form the parent ($J/\psi$, $\pi$) mesons, which are therefore automatically in a planar configuration.}.
  We consider the 5--parton tree--level amplitudes, $gg \to ggg$, $q\overline{q}  \to ggg$ and $q\overline{q}  \to q\overline{q} g$ (including other initial/final--state crossings), and show that in all cases planar zeros exist for particular colour configurations. This is derived using the MHV formalism~\cite{Mangano:1990by}, which allows simple expressions for all non--zero 5--parton QCD amplitudes to be written down at tree--level. In addition, the  `BCJ' relations~\cite{Bern:2008qj}, which allow the $n$--parton QCD amplitudes to be written as a linear combination of only $(n-3)!$ independent partial amplitudes, play an important role in these results. These relations rely upon a new identity between the kinematic terms for particular diagrams, analogous to the Jacobi identity satisfied by the colour factors. As noted in~\cite{Bern:2008qj}, this generalises an identity for the 4--parton amplitude that has already been derived and used in~\cite{Zhu:1980sz,Goebel:1980es} to explain the presence of the type--I radiation zeros discussed above. Here, we will see that the result of~\cite{Bern:2008qj} permits a very simple derivation of the planar condition necessary for type--II radiation zeros in these 5--parton QCD amplitudes. We will then make use of the MHV formalism to show that the remaining zero condition, although occurring from a delicate cancellation of individual Feynman diagrams, can in fact be written as a simple relation involving only rapidity differences of the final--state partons. Although conditions in the case of QED processes with soft photon emission have been derived in~\cite{Heyssler:1997fy,Stirling:1999sj,Rodriguez:2003tt}, these results represent the first analytic expressions for the existence of such planar zeros for general particle momenta. 
  
While these colour--dependent effects may generally be of only theoretical interest, being washed out by the usual colour--averaging in inclusive production, a zero is also found to exist when the incoming gluons are in a colour--singlet state, both for the 5--gluon and, in certain circumstances, $gg \to q\overline{q} g$ amplitudes. These are precisely the amplitudes which contribute to central exclusive  three--jet  production: in this process only the system $X=jjj$ and no additional activity is produced in the central region, while the colliding protons remain intact and scatter at near zero angle to the beam line (see~\cite{Albrow:2010yb,Harland-Lang:2014lxa} for reviews and further references, as well as discussion of how the exclusive signal can be defined in the case of jet production). Crucially, such a final--state requires the incoming gluons in the $gg \to ggg$ and $gg \to q\overline{q} g$ production subprocesses to be in a colour--singlet state; these planar radiation zeros should therefore be in principle observable in this process. A more detailed phenomenological study of this possibility is planned in~\cite{HKRfut}, while here we focus on the general aspects of these zeros.

Finally, although the main emphasis of this paper is on the planar zeros discussed above, we also demonstrate the existence of zeros  for non--planar parton configurations, but which on the other hand do not generally satisfy the simple condition (\ref{type1}), and are therefore neither of type--I or II. We derive explicit analytic conditions for the presence of such zeros for a subset of the 5--parton QCD scattering processes. 

The outline of this paper is as follows. In Section~\ref{sec:prelim} we present a summary of the MHV formalism. In Section~\ref{sec:presence} we demonstrate the existence of planar zeros at tree--level in the 5--parton QCD amplitudes and give concise analytic expressions for the corresponding zero conditions: in Section~\ref{firstlook} the simplified case of the 5--gluon amplitude with colour--singlet initial--state gluons is considered; in Section~\ref{secgenc} this is generalised to the case of arbitrary gluon colour; in Section~\ref{sec:quarks} the quark--mediated $gg\to q\overline{q}g$ and $qq \to qq g$ processes  (including other initial/finial--state crossings) are considered. In Section~\ref{results} representative numerical results are presented: in Section~\ref{res:planar} the planar zeros are considered; in Section~\ref{secnonplanar} non--planar zeros are considered. In Section~\ref{conclusions} a summary and outlook is presented. Finally, some useful MHV formulae as well as explicit expressions for the non--planar zero conditions and the colour--summed 5--gluon and $q\overline{q} \to ggg$ squared matrix elements are given in appendices.

\section{MHV formalism}\label{sec:prelim}

It is well known (see for example~\cite{Mangano:1990by}) that the tree--level $n$--gluon scattering amplitudes in which $n$ or $n-1$ gluons have the same helicity vanish completely, while those with ($n-2$) same--helicity gluons, the so--called `maximally helicity violating' (MHV), or `Parke--Taylor', amplitudes, are given by remarkably simple formulae~\cite{Parke:1986gb,Berends:1987me}. These results were extended using supersymmetric Ward identities to include amplitudes with one and two quark--antiquark pairs in~\cite{Mangano:1990by}, where `MHV' refers to the case where ($n-2$) partons have the same helicity. In these cases, simple analytic expressions can again be written down for the MHV amplitudes, while for greater than 2 fermion--anti--fermion pairs (recalling that the helicities of a connected massless fermion--anti--fermion pair must be opposite) no MHV amplitudes exist.

In this framework, the full $n$--parton scattering amplitude $\mathcal{M}_n$ can be written in the form of a `dual expansion', as a sum of products of colour factors $T_n$ and purely kinematic partial amplitudes $A_n$
\begin{equation}\label{mhv}
\mathcal{M}_n(\{k_i,h_i,c_i\})=ig^{n-2}\sum_\sigma T_n(\sigma\{c_i\})A_n(\sigma\{1^{\lambda_1},\cdots,n^{\lambda_n}\})\;,
\end{equation}
where $c_i$ are colour labels, $i^{\lambda_i}$ corresponds to the $i$th particle ($i=1\cdots n$), with momentum $k_i$ and helicity $\lambda_i$, and the sum is over appropriate simultaneous non--cyclic permutations $\sigma$ of colour labels and kinematics variables. The purely kinematic part of the amplitude $A_n$ encodes all the non--trivial information about the full amplitude, $\mathcal{M}_n$, while the $T_n$ are given by known colour factors.

Considering the $n$--gluon and $q\overline{q}$ + $(n-2)$ gluon partial amplitudes, these are given by
\begin{align}\label{mhvpartg}
A_n(g_1^+,...\,,g_i^-,...\,,g_j^-,...\,,g_n^+)&=\frac{\langle i\, j \rangle^4}{\prod_{k=1}^n \langle k\, k+1 \rangle}\;,\\ \label{mhvpartq}
A_n(q_{1}^+,...\,,g_i^-,...\,,\overline{q}_n\!\!{}^-)&=\frac{\langle i\, n \rangle^3\langle i\,1 \rangle}{\prod_{k=1}^n \langle k\, k+1 \rangle}\;,
\end{align}
where the `$...$' indicate the emission of an arbitrary number of positive helicity gluons, $\langle k_i\,k_j\rangle \equiv \langle k_i^-|k_j^+\rangle = \overline{u}_-(k_i)u_+(k_j)=\overline{v}_+(k_i)v_-(k_j)$ is the standard spinor contraction, and all momenta are defined as incoming. In the case of the quark amplitudes, we can see that additional gluons can only be emitted from one side of the quark line, reducing the number of contributing partial amplitudes. The equivalent `$\overline{{\rm MHV}}$' amplitudes with $+\leftrightarrow -$ are related trivially to the above expressions.

The corresponding colour factors $T_n$ are given by
\begin{align}\label{colourg}
 T_n(g_1,...\,,g_n)&={\rm Tr}(\lambda^{c_1}...\,\lambda^{c_n})\;,\\ \label{colourq}
 T_n(q_1,g_2,...\,,g_{n-1},\overline{q}_n)&=(\lambda^{c_2}...\,\lambda^{c_{n-1}})_{i_1}^{\;\:i_n}\;, 
\end{align}
where $c_n$ is the colour index of the $n$--th gluon, $i$ the colour index of the quark/anti--quark, and the $\lambda^{c_n}$ are the usual Gell--Mann matrices, normalized as in~\cite{Mangano:1990by}. 

From (\ref{mhvpartq}) it follows that in the $q\overline{q}$ + $(n-2)$ gluon case there are in general $(n-2)!$ independent kinematic amplitudes. In the $n$--gluon case, the apparently $(n-1)!$ independent amplitudes can in fact be reduced to $(n-2)!$~\cite{DelDuca:1999ha,DelDuca:1999rs}, by making use of known (so--called `Kleiss--Kuijf'~\cite{Kleiss:1988ne}) relations between the partial amplitudes, with
\begin{equation}\label{glukk}
\mathcal{M}_n(g_1,...\,,g_n)=g^{n-2}\sum_\sigma (T^{c_{\sigma_2}}...\,T^{c_{\sigma_{n-2}}})_{c_1 c_n} A_n(1,\{\sigma_2,...\,,\sigma_{n-1}\},n)\;,
\end{equation}
where the sum is over all permutations of the $(n-2)$ elements, with the positions of gluons 1 and $n$ held fixed; the amplitudes $A_n$ are given as before by (\ref{mhvpartg}). The $T_a$ are the $SU(N)$ generators in the adjoint representation, related to the usual QCD structure constants. It is interesting to compare (\ref{glukk}) with (\ref{mhvpartq}) and (\ref{colourq}): the only difference between the $n$--gluon and $q\overline{q}$ + $(n-2)$ gluon amplitudes is in the representation of the colour matrices, i.e. adjoint in the gluon case and fundamental for the $q\overline{q}$.

In fact, this number of independent amplitudes can be further reduced using the so--called `BCJ' relations~\cite{Bern:2008qj}. These employ an identity between the kinematic terms for particular diagrams, which is analogous to the Jacobi identity obeyed by the color factors, to obtain non--trivial relations between the partial amplitudes. This allows the $n$--parton amplitudes to be written as a linear combination of only $(n-3)!$ independent partial amplitudes; for the $5$--parton case we will consider here, this implies that there are only two independent partial amplitudes, with all others expressed in terms of this linear basis.

Finally, for the amplitudes with two $q\overline{q}$ pairs, the colour factor is given by
\begin{equation}\label{colourqq}
T_n(q_1,...\,,\overline{q}_j,q_{j+1},...\,,\overline{q}_n)=\frac{(-1)^p}{N_C^p}(\lambda^{c_2}...\,\lambda^{c_{j-1}})_{i_1}^{\;\:i_j}(\lambda^{c_{j+2}}...\,\lambda^{c_{n-1}})_{i_{j+1}}^{\;\:i_n}\;,
\end{equation}
where $p=0$ (1) when $q_1$ is connected by a fermion line to $\overline{q}_j$ ($\overline{q}_n$); these correspond to the leading (subleading) colour terms. The total amplitude is then given by summing over all partitions of the $(n-4)$ gluons between the $q\overline{q}$ pairs, and over permutations of the gluon indices. If no gluons are emitted between a given $q\overline{q}$ pair, then the corresponding product of Gell--Mann matrices is replaced by a Kronecker delta. The kinematic partial amplitudes for the leading colour term is given by
\begin{equation}\label{mhvpartqq}
 A_n(q_1^{h_1},...\,,\overline{q}_j^{\;-h_2},q_{j+1}^{h_2},...\,,\overline{q}_n^{\;-h_1})=\frac{F(h_1,h_2)\langle 1\, j \rangle \langle n\, j+1 \rangle}{\prod_{k=1}^n \langle k\, k+1 \rangle}\;,
\end{equation}
where the prefactor $F(h_1,h_2)$ depends on the quark helicities, $h_i$, see e.g.~\cite{Birthwright:2005vi} for explicit expressions, and for the subleading term we replace $\overline{q}_j \leftrightarrow \overline{q}_n$ in the particle ordering and $j \leftrightarrow n$ in the numerator. If the $q\overline{q}$ pairs are identical then the amplitudes with the quarks interchanged that are consistent with helicity conservation along the quark lines should be included (with an overall minus sign for the odd permutation).



A range of useful identities satisfied by the spinor contractions, and other explicit expressions which will be used in this paper are given in Appendix~\ref{app:spinor}.

\section{Presence of zeros}\label{sec:presence}

For the 5--parton scattering processes we will consider in this paper, the only non--zero amplitudes are MHV, and so we can make use of this formalism throughout. 
The only difference between the amplitudes (\ref{mhvpartg},\ref{mhvpartq}) and (\ref{mhvpartqq}) for each parton helicity configuration is in the numerator, which will factor out in the total amplitude (\ref{mhv}). The presence of a radiation zero will therefore rely upon a cancelation between the denominator terms in the partial amplitudes, summed over the particle orderings. As this is independent of the particular particle helicities, we can safely ignore these, and the corresponding numerator terms, in what follows.

We will consider the amplitudes for the $5$--parton scattering process
\begin{equation}\label{ggg}
 P^{a_1}(k_1)+P^{a_2}(k_2)\to P^{a_3}(k_3)+P^{a_4}(k_4)+P^{a_5}(k_5)\;,
\end{equation}
where $P=q,\overline{q},g$, the $k_j$ are the particle four--momenta, and $a_j=c_j (i_j)$ are the colour indices, in the corresponding adjoint (fundamental) representation.

Before considering the more general case we will show how a radiation zero arises in the simpler example of the 5--gluon amplitude with the incoming gluons in a colour--singlet state.

\subsection{First look: colour--singlet $5$--gluon amplitude}\label{firstlook}

Considering the $n=5$ gluon amplitude, we can make use of the decomposition (\ref{glukk}), with the gluons in the $i=1,5$ positions corresponding to momenta $k_1$, $k_2$ above, i.e.
\begin{equation}\label{glukkex}
\mathcal{M}_5=g^3\left( (T^{c_3}T^{c_4}T^{c_5})_{c_1 c_2}\, A_5(1,3,4,5,2)+{\rm permutations}\right)\;.
\end{equation}
While, as will see in Section~\ref{secgenc}, the results below can be derived simply using the BCJ relations, we will first consider this more explicit decomposition for the sake of demonstration. When gluons 1 and 2 are in a colour--singlet configuration, we are interested in the case that
\begin{equation}\label{m5}
\mathcal{M}_5^{\rm cs} \sim f^{c_3 c_4 c_5}\left(A_{345}-A_{354}-A_{435}+A_{453}-A_{543}+A_{534}\right)=0\;,
\end{equation}
where
\begin{equation}\label{adef}
 A_{ijk}\equiv \frac{1}{\langle 1\, 2 \rangle\langle 2\, i \rangle\langle i\, j \rangle\langle j\, k \rangle\langle k\, 1 \rangle}\;,
\end{equation}
is the kinematic partial amplitude for a given ordering of the final--state gluons, with the numerator factored out. We can see that the colour coefficients completely factorize, so that if the kinematic term inside the brackets in (\ref{m5}) vanishes, then a zero will occur. To explore when or if this can happen, we can first make use of the Schouten identity (\ref{sc}) to simplify this expression, giving the relatively simple result
\begin{equation}\label{zero}
 \langle 3\, 5 \rangle^2\langle 1\, 4 \rangle \langle 2\, 4 \rangle +\langle 3\, 4 \rangle^2 \langle 1\, 5 \rangle \langle 2\, 5 \rangle + \langle 4\, 5 \rangle^2\langle 1\, 3 \rangle \langle 2\, 3 \rangle = 0\;.
\end{equation}
We can then make further use of the Schouten identity (\ref{sc}), as well as the constraint of $4$--momentum conservation (\ref{cons}), to show that (\ref{zero}) is equivalent to
\begin{equation}\label{zero1}
f_{45}\frac{\langle 1\, 4\rangle\langle 2\, 5\rangle}{\langle 1\, 5\rangle\langle 2\, 4\rangle}+f_{54}\frac{\langle 1\, 5\rangle\langle 2\, 4\rangle}{\langle 1\, 4\rangle\langle 2\, 5\rangle}+f_{45}+f_{54}-8=0\;,
\end{equation}
where
\begin{equation}
 f_{45}=1-\frac{s_{24}}{s_{23}}-\frac{s_{15}}{s_{13}}\;,
\end{equation}
and similarly for $f_{54}$, for which we exchange $4 \leftrightarrow 5$. We note that the spinor contractions $\langle i\, j \rangle$ are complex functions of the four--momenta $k_{i},k_{j}$, and therefore this relation in fact corresponds to two separate conditions on the particle momenta; we will see this explicitly below. By considering the complex conjugate of (\ref{zero1}) and combining these two relations we find that they imply
\begin{equation}\label{csq}
 \left(\frac{\langle 1\, 4\rangle\langle 2\, 5\rangle}{\langle 1\, 5\rangle\langle 2\, 4\rangle}\right)^2 = \frac{s_{14}s_{25}}{s_{15}s_{24}}\;,
\end{equation}
that is, this ratio must be purely real. This result assumes that $f_{45}\neq \frac{s_{15}s_{24}}{s_{14}s_{25}}f_{54}$: we will comment on this additional solution below. There is nothing special about this choice of gluon momenta in (\ref{csq}), and more generally we have
\begin{equation}\label{cij}
 \left(\frac{\langle 1\, i\rangle\langle 2\, j\rangle}{\langle 1\, j\rangle\langle 2\, i\rangle}\right)^2 = \frac{s_{1i}s_{2j}}{s_{1j}s_{2i}}\;,
\end{equation}
for some choice of $i,j=3,4,5$, with $i\neq j$.

What does this condition imply? To clarify this we can consider the explicit representation (\ref{cont}) for the spinor contractions given in Appendix~\ref{app:spinor}, which gives\footnote{In fact the contraction $\langle 2\, j \rangle$ is not well defined in the basis we have chosen here, as it depends on the phase $\exp(i\phi_2)$, which is not specified for a particle moving along the $z$--axis. However this dependence cancels in the ratio (\ref{cij}). Alternatively a different basis that does not suffer from this issue may be chosen, for example the Weyl representation of the $\gamma$ matrices, with the same result in the end.}
\begin{equation}
\left(\frac{\langle 1\, i\rangle\langle 2\, j\rangle}{\langle 1\, j\rangle\langle 2\, i\rangle}\right)^2=\frac{s_{1i}s_{2j}}{s_{1j}s_{2i}}\,\cdot\,e^{2i\Delta\phi_{ij}}\;,
\end{equation}
for general momenta $k_i$, $k_j$, where $\Delta\phi_{ij}=\phi_i-\phi_j$ is the difference in azimuthal angle between the particles. Comparing this with (\ref{cij}) we can see that this corresponds to the condition
\begin{equation}\label{phireq}
 \sin \Delta\phi_{ij}=0\;,
\end{equation}
that is, the outgoing gluons are either aligned ($\phi_i=\phi_j$) or anti--alligned ($\phi_i=\phi_j+\pi$) in the transverse direction\footnote{For the $3$--gluon final state we are considering here, we must of course have two alligned gluons and one anti--aligned gluon for momentum to be conserved, e.g. $\phi_1=\phi_2$, $\phi_3=\phi_{1,2}+\pi$.}. In other words,  {\it all five gluon momenta must lie in a plane}. This is precisely the condition required in the case of type--II radiation zeros that was anticipated in the introduction.

However, this first condition is necessary, but not sufficient, for the presence of a zero. Applying the constraint (\ref{phireq}) to (\ref{zero}) and making use of (\ref{prodex}), we arrive at an expression for the additional condition which must be satisfied for a zero to occur:
\begin{equation} \label{master1}
 {\rm sinh}^2\frac{\Delta_{34}}{2}={\rm cosh}^2\frac{\Delta_{45}}{2}+{\rm cosh}^2\frac{\Delta_{35}}{2}\;,
\end{equation}
where $\Delta_{ij}=y_i-y_j$ is the rapidity difference between the gluons, as in (\ref{prodex}), and we have defined, without loss of generality, gluons $3$ and $4$ to be aligned (i.e. $\phi_3=\phi_4$). It is interesting to observe the relative simplicity of this expression, which is written purely in terms of  rapidity differences of the final--state partons, of which two are independent.
 

\subsection{$5$--gluon amplitudes: general colour}\label{secgenc}

The planar condition (\ref{phireq}) can in fact be most readily shown to follow from the BCJ conditions described in Section~\ref{sec:prelim}; this is particularly useful for the more involved case of general colour. These relations express the 6 apparently independent amplitudes in (\ref{m5}) in terms of two basis amplitudes, $A_{534}$ and $A_{435}$, say. In this case we have for example~\cite{Bern:2008qj}
\begin{equation}\label{bcj}
A_{345}=\frac{s_{15}s_{24}A_{534}+s_{14}(s_{45}-s_{25})A_{435}}{s_{13}s_{45}}\;,
\end{equation}
while similar expressions can be written down for the other 3 amplitudes, $A_{543}$, $A_{453}$ and $A_{354}$ (in all cases defined as in (\ref{adef})). Using these results, the zero condition can be written for general colour as
\begin{equation}\label{zerog1}
N_1({k_i,c_i})A_{534}+N_2({k_i,c_i})A_{435}=0\;,
\end{equation}
where $N_{1,2}$ are \emph{real} functions of the parton 4--momenta and colours, $k_i$ and $c_i$; we do not give explicit expressions for these here for the sake of brevity. Rearranging, and substituting explicit expressions for $A_{534}$ and $A_{435}$, we find this implies that
\begin{equation}\label{zerog}
\frac{\langle 1\, 4\rangle\langle 2\, 5\rangle}{\langle 1\, 5\rangle\langle 2\, 4\rangle}=-\frac{N_1({k_i,c_i})}{N_2({k_i,c_i})}\;,
\end{equation}
in other words the same ratio as in (\ref{csq}) for the colour--singlet case must be purely real. Thus, for general gluon colour, we have the same constraint that the gluon four--momenta must lie in a plane. We note that this result can also be derived using the more involved procedure of the previous section. Indeed, for the case of MHV amplitudes, the BCJ relations are in fact direct results of the momentum conservation relations (\ref{cons}), as well as the Schouten identity (\ref{sc}) (for non--MHV amplitudes, where these relations also hold, this is of course no longer the case). 

The relation (\ref{zerog}), subject to this requirement, then defines the second kinematic condition. After some algebra, we find that  (\ref{master1}) generalises to
\begin{align}\nonumber
&C_{45}\,\rm{sinh} \Delta_{45}-C_{35}\,{\rm sinh} \Delta_{35}-C_{34}\,{\rm sinh} \Delta_{34} +\\ \label{master1g}
&2\left(\tilde{C}_{45}\,{\rm cosh}^2\frac{\Delta_{45}}{2}-\tilde{C}_{35} \,{\rm cosh}^2\frac{\Delta_{35}}{2}-\tilde{C}_{34} \,{\rm sinh}^2\frac{\Delta_{34}}{2}\right)=0\;,
\end{align}
where we again assume without loss of generality that gluons $3$ and $4$ are aligned, while
\begin{align}\label{c45}
C_{45}&=(T^{c_4} T^{c_3} T^{c_5}+T^{c_5} T^{c_3} T^{c_4})_{c_1 c_2}\;,\\ \label{ct45}
\tilde{C}_{45}&=(T^{c_4} T^{c_3} T^{c_5}-T^{c_5} T^{c_3} T^{c_4})_{c_1 c_2} \;,
\end{align}
and similarly for $C_{34}$, $C_{35}$, after interchanging the corresponding colour indices. This expression again only depends on rapidity differences between the final--state gluons, as in (\ref{master1}), but now contains an explicit colour dependence. The colour--singlet identification in Section~\ref{firstlook} simply corresponds to taking the traces of (\ref{c45}) and (\ref{ct45}), in which case we readily arrive back at (\ref{master1}). The existence and form of any such radiation zero will in general depend on the colour indices of the gluons, as we would expect. 

Finally, we can see that (\ref{zerog1}) will also be satisfied if we have
\begin{equation}\label{otherzero}
N_1({k_i,c_i})=N_2({k_i,c_i})=0\;.
\end{equation}
An equivalent possibility for the colour--singlet case was observed below (\ref{csq}), see footnote 3. These again correspond to two constraints on the gluon momenta (for a given colour configuration), as in the planar case, but with in general no requirement that the gluon momenta should lie in a plane. We might expect these to be satisfied for some region of phase space, and thus for such `non--planar' zeros to occur.
 We will postpone further discussion of this to Section~\ref{secnonplanar}, where we we show that (\ref{otherzero}) can indeed be satisfied for certain colour configurations.
 
\subsection{Amplitudes with quarks}\label{sec:quarks}

Having discussed the purely gluonic case, it is natural to ask whether these results extend to amplitudes including $q\overline{q}$ pairs. As discussed in Section~\ref{sec:prelim}, the expressions (\ref{mhvpartq}) and  (\ref{glukk}) for the  $q\overline{q}$ + $(n-2)$ gluon and  $n$--gluon amplitudes only differ in the representation of the colour matrices. For the $q\overline{q}\to ggg$ amplitude
\begin{equation}
q^{i_1}(k_1) \,\overline{q}^{\,i_2} (k_2)\to g^{c_3}(k_3)\,g^{c_4}(k_4)\,g^{c_5}(k_5)\;,
\end{equation}
 this implies that exactly the same zero conditions as (\ref{master1g}) hold, but with e.g.
\begin{align}\label{c45q}
C_{45}&\to(\lambda^{c_4} \lambda^{c_3} \lambda^{c_5}+\lambda^{c_5} \lambda^{c_3} \lambda^{c_4})_{i_1}^{\;\:i_2}\;,\\ \label{ct45q}
\tilde{C}_{45}&\to(\lambda^{c_4} \lambda^{c_3} \lambda^{c_5}-\lambda^{c_5} \lambda^{c_3} \lambda^{c_4})_{i_1}^{\;\:i_2} \;.
\end{align}
We may therefore expect similar zeros in the case of these amplitudes.  For illustration, if as in Section~\ref{firstlook} we consider the simplifying scenario that the initial--state partons (in this case quarks) are in a colour--singlet configuration, we have
\begin{align}\label{c45qs}
C_{45}&\to\frac{1}{2}d^{c_4 c_3 c_5}\;,\\ \label{ct45qs}
\tilde{C}_{45}&\to \frac{i}{2} f^{c_4 c_3 c_5}\;,
\end{align}
where $f^{c_4 c_3 c_5}$ ($d^{c_4 c_3 c_5}$) are the usual anti--symmetric (symmetric) structure constants, and equivalent results hold for $C_{34}$, $C_{35}$. The zero condition will therefore depend on the specific final--state colour configuration: for colour assignments where the $f^{c_3 c_4 c_5}$ are non--zero the condition is exactly as in the section~\ref{firstlook} for the all-gluon colour--singlet amplitude, while for the case that the $d^{c_3 c_4 c_5}$ are non--zero a distinct condition exists. On inspection it is found that the latter condition in fact admits no solutions in the physical phase space region. We note that there is no final--state colour configuration for which both  $f^{c_3 c_4 c_5}$ and $d^{c_3 c_4 c_5}$ are non--zero. More generally, the colour coefficients (\ref{colourq}) are all either purely real or imaginary, depending on the colour assignment. This implies that the equivalent condition to (\ref{zerog}) for amplitudes with quarks corresponds to the same planar condition as in the purely gluonic case, i.e. the right hand side of  (\ref{zerog}) is purely real.

For the remaining $q\overline{q}ggg$ scattering processes, it is sufficient to consider
\begin{align}\label{qgq}
q^{i_1}(k_1) \,g^{c_2}(k_2)&\to q^{i_3}(k_3)\,g^{c_4}(k_4)\,g^{c_5}(k_5)\;,\\ \label{ggq}
g^{c_1}(k_1)\, g^{c_2}(k_2)&\to g^{c_3}(k_3)\,q^{i_4}(k_4)\,\overline{q}^{\,i_5} (k_5)\;,
\end{align}
with the results for the $\overline{q}g$--initiated process being identical to the $qg$--initiated.
There is no simple relation to the condition (\ref{master1g}), since the quark (anti--quarks) which correspond to the $i=1,5$ positions in (\ref{mhvpartq}) no longer carry the momenta $k_1$, $k_2$, as in the decomposition of (\ref{glukkex}). However, again results for the zero conditions can be found in a similar fashion to those described above; indeed, by suitably interchanging the particle momenta and colour labels,  it is possible to derive these fairly trivially. While the planar condition remains, the expressions  for the additional constraint are no longer quite as simple as (\ref{master1g}). For example, in the $gg \to g q\overline{q}$ case we have
\begin{align}\nonumber
&8\,{\rm cosh}\frac{\Delta_{35}}{2}{\rm sinh}\frac{\Delta_{34}}{2}\left(C_{12}\,{\rm sinh}\frac{\Delta_{45}}{2}+\tilde{C}_{12}\,{\rm cosh}\frac{\Delta_{45}}{2}\right)+ e^{\Delta_{35}}(\tilde{C}_{23}-C_{23})\\ \label{masterq}
&+e^{\Delta_{34}}(\tilde{C}_{23}+C_{23})
+ e^{-\Delta_{35}}(\tilde{C}_{13}-C_{13})+e^{-\Delta_{34}}(\tilde{C}_{13}+C_{13})-2(C_{13}+C_{23})=0\;,
\end{align}
for the case that the gluon and quark in the final--state are aligned and  
\begin{align}\label{c12}
C_{12}&=(\lambda^{c_1} \lambda^{c_3} \lambda^{c_2}+\lambda^{c_2} \lambda^{c_3} \lambda^{c_1})_{i_4}^{\;\:i_5}\;,\\ \label{ct12}
\tilde{C}_{12}&=(\lambda^{c_1} \lambda^{c_3} \lambda^{c_2}-\lambda^{c_2} \lambda^{c_3} \lambda^{c_1})_{i_4}^{\;\:i_5} \;.
\end{align}
A similar, but not identical, constraint can be written down for the case that the $q\overline{q}$ pair are aligned, although we omit this for the sake of brevity here. In the $qg \to q gg$ case, the resulting constraints are a little more involved, but can still readily be written down; again these are omitted for the sake of brevity. In all cases, we have the requirement that the momenta of the five partons must lie in a plane, combined with an additional constraint which can be expressed purely in terms of rapidity differences of the final--state partons. If we consider (\ref{masterq}) for the case of colour--singlet initial--state gluons we find
\begin{equation}\label{condq}
{\rm cosh}\frac{\Delta_{35}}{2}{\rm sinh}\frac{\Delta_{34}}{2}\,{\rm sinh}\frac{\Delta_{45}}{2}+4\,{\rm cosh}^2\frac{\Delta_{35}}{2}-4\,{\rm sinh}^2\frac{\Delta_{34}}{2}=0\;,
\end{equation}
while if the $q\overline{q}$ pair are aligned we have
\begin{equation}\label{condq1}
{\rm cosh}\frac{\Delta_{35}}{2}{\rm cosh}\frac{\Delta_{34}}{2}\,{\rm cosh}\frac{\Delta_{45}}{2}+4\,{\rm cosh}^2\frac{\Delta_{35}}{2}+4\,{\rm cosh}^2\frac{\Delta_{34}}{2}=0\;,
\end{equation}
where in both cases we have set $N_c=3$. While the latter condition clearly admits no solution, we will see in Section~\ref{res:planar} that the former does.

Finally, we can also consider the $q\overline{q}q\overline{q} g$ processes, e.g,
\begin{align}\label{4q}
q^{i_1}(k_1) \,q^{i_2}(k_2)&\to g^{c_3}(k_3)\,q^{i_4}(k_4)\,q^{\,i_5} (k_5)\;,\\ \label{4q1}
q^{i_1}(k_1) \,g^{c_2}(k_2)&\to q^{i_3}(k_3)\,\overline{q}^{i_4}(k_4)\,q^{i_5}(k_5)\;,
\end{align}
and similarly for other quark/anti--quark interchanges.
While a similar logic to that described above may be followed to determine the corresponding zero conditions,  if the scattered quarks are identical the situation is somewhat more complicated, as the contributing amplitudes now depend on the quark helicity configurations. For example, considering the process (\ref{4q}), if, following the labelling of (\ref{mhvpartqq}), we have $(h_1,h_2)=(+,+)$, then the $t$ and $u$--channel amplitudes (with the latter corresponding to quark interchange) must be included and will interfere, while for the $(h_1,h_2)=(+,-)$ configuration, helicity conservation along the quark lines implies that either the $t$ or the $u$ channel amplitudes contribute, depending on the helicities of the final--state. Thus, if a radiation zero is to be present, \emph{both} $u$ and $t$--channel contributions must individually vanish. However, upon inspection it is found that this does indeed occur for some colour configurations (for example, the vanishing of the $t$--channel contribution may imply the vanishing of the $u$--channel contribution, and vice versa) and so zeros are present. For non--identical quarks, where only the $t$--channel amplitudes contribute, zeros occur in these and in a wider range of cases.


\section{Results and discussion}\label{results}

\subsection{Planar zeros}\label{res:planar}
 
In this section we explore whether the analytic expressions given above in fact contain solutions in the physical region, and therefore correspond to genuine radiation zeros.
 
To explore what these constraints imply for this $2\to 3$ process, we recall that we have in general 12 unknown 4--momentum components of the outgoing partons, with 4 constraints from energy--momentum conservation, 3 from the parton on--shell conditions and 1 from the requirement that the reaction occurs in a plane\footnote{More precisely, the condition corresponds to requiring that $\Delta\phi_{ij}=0$ for one choice of pairings $i,j$; the fact that the reaction should be planar then comes from momentum conservation.}, leaving in principle 4 unknowns. However one of these corresponds to the overall orientation of the reaction plane, which has no physical consequence. We are therefore left with 3 unknowns, and so the three parton rapidities $y_{i}$ can be conveniently used to define the final state. Thus we have a three--dimensional phase space volume with coordinates $(y_3,y_4,y_5)$, and the zero conditions will in general define a two--dimensional surface within this. As these conditions only depend on rapidity differences, if we instead consider $\Delta_{35}$ and $\Delta_{45}$, say, the these will define a curve in the $\Delta_{35}-\Delta_{45}$ plane. 

 \begin{figure}
 \begin{center}
 \includegraphics[scale=0.55]{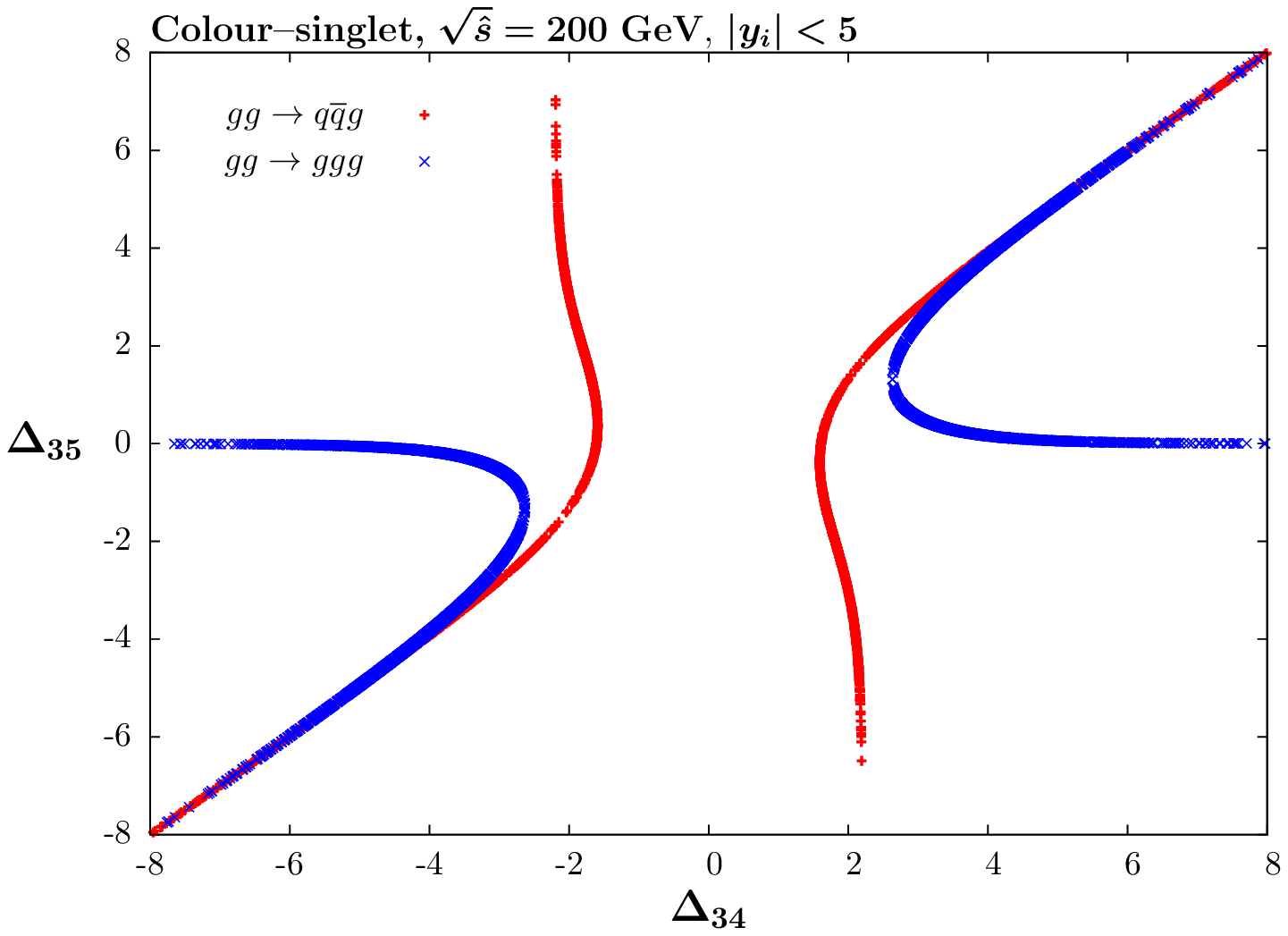}
  \includegraphics[scale=0.55]{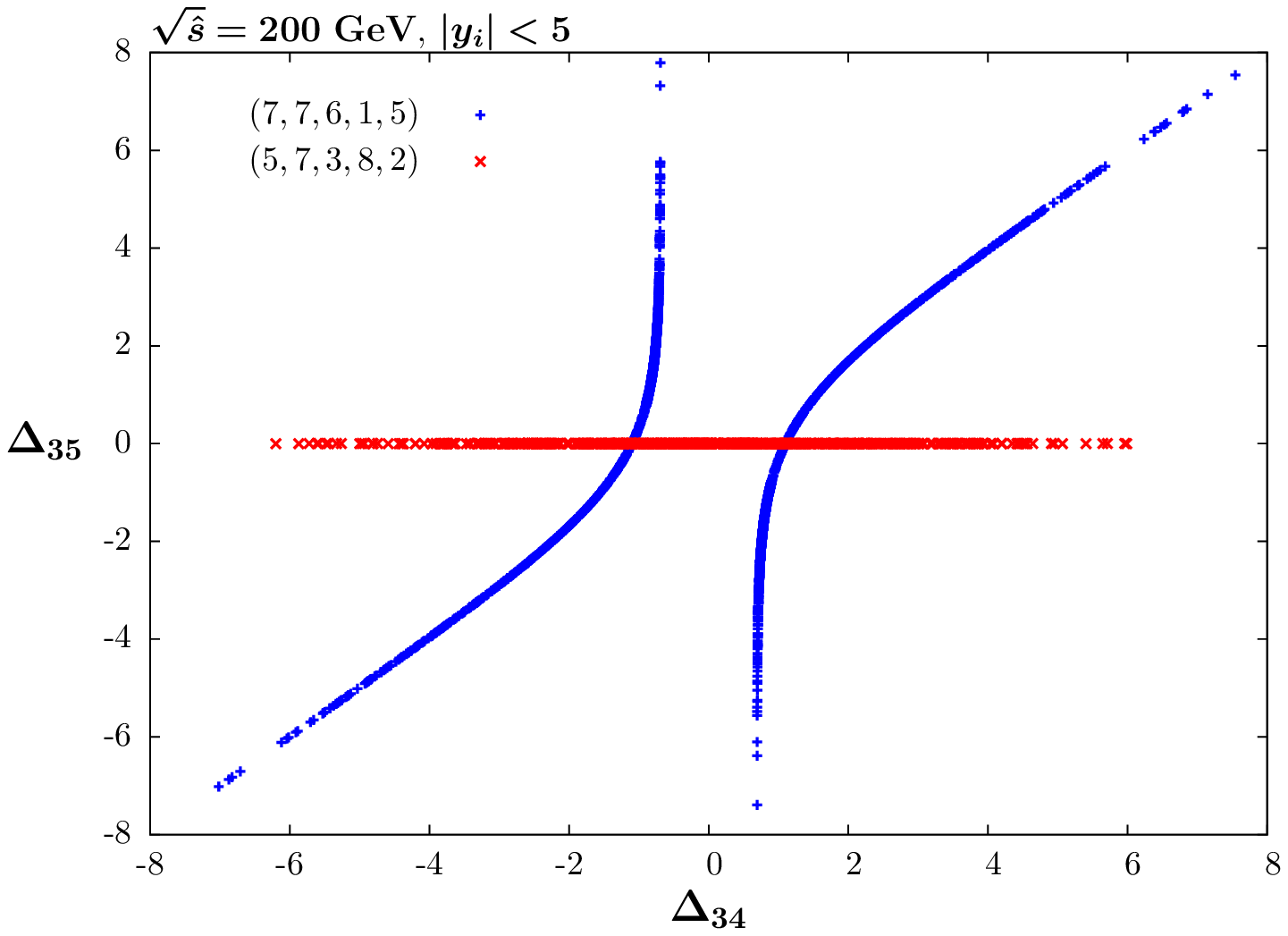}
 \caption{Scan of solutions to (\ref{master1g}) and (\ref{masterq}) which lie in the physical region. 
(Left) Solutions for the $gg\to ggg$ and $gg\to q\overline{q} g$ amplitudes for which the incoming gluons are in a colour--singlet state. For the quark amplitude, the case where the gluon and quark are aligned (with momenta defined as in (\ref{ggq})) is shown; if the gluon and anti--quark are aligned the corresponding curve is simply reflected around the line $\Delta_{34}=\Delta_{45}$. (Right) Solutions for the 5--gluon amplitudes, for two different colour choices $(c_1,c_2,c_3,c_4,c_5)$. The central system mass is $\sqrt{\hat{s}}=200$ GeV, and the final--state particles are required to have rapidity $|y_i|<5$.
 }\label{y34scan}
 \end{center}
 \end{figure}

 To verify whether these zeros occur in the physical region, we perform a scan over the allowed $\Delta_{35}$--$\Delta_{45}$ region and determine numerically those values for which the conditions, e.g. (\ref{master1g}), hold in the physically allowed region (i.e. subject to $4$--momentum conservation and all final--state particles having positive energy). We take $\sqrt{\hat{s}}= 200$ GeV, and require the final--state gluon rapidities to satisfy $|y_i|<5$, although the results do not depend sensitively on this choice. For demonstration purposes no further cuts on the parton separation or transverse momentum $k_\perp$ are imposed; the effect of these is simply to limit the region of allowed solutions. 
 
 Some selected results are shown in Fig.~\ref{y34scan}. These plots are shown for demonstration purposes only: the density of points depends on the precise scan procedure and does not have any direct physical relevance. Fig.~\ref{y34scan} (left) corresponds to the case of colour--singlet initial--state gluons for the 5--gluon\footnote{In addition, as discussed in Section~\ref{sec:quarks}, the solution for the 5--gluon amplitude corresponds to the $q\overline{q}\to ggg$ amplitude, for certain colour configurations.} and $gg\to q\overline{q}g$ amplitudes (in the latter case where the quark and gluon are aligned): we can see, as expected from the discussion above, that the zero conditions define a curve in the $\Delta_{35}$--$\Delta_{45}$ plane (or two curves, which are identical after suitable transformations, for cases where a $y_i \to -y_i$ symmetry is exhibited in the conditions). As discussed in Section~\ref{sec:quarks}, if the $q\overline{q}$ pair are aligned then no zero occurs. It is interesting to note that in the high $|\Delta_{34}|, |\Delta_{35}|$ region, and when $\Delta_{45}\approx 0$, the zeros in the 5--gluon and $gg \to q\overline{q}g$ amplitudes coincide, occurring for $\Delta_{34}\approx\Delta_{35}$; this can readily be confirmed analytically by comparing (\ref{master1}) and (\ref{condq}). We also show in Fig.~\ref{y34scan} (right) the corresponding zero curves for the 5--gluon amplitude only, for two representative choices of general colour indices, $c_i$, defined as in (\ref{ggg}). Although such specific colour choices clearly do not correspond to physically observable situations, this nonetheless demonstrates that the colour--singlet condition shown in  Fig.~\ref{y34scan} (left) does not correspond to a special case. Indeed, for a wide range of colour configurations similar curves for which the zero conditions are satisfied are found to occur. Moreover,  the case of all of the $q\overline{q}ggg$ and $q\overline{q}q\overline{q}g$ amplitudes has also been examined numerically, and again similar zero curves are found to occur for certain colour configurations. We do not show these explicitly here for the sake of brevity, and as the precise shape of the specific curves does not provide any further insight.

 \begin{figure}
\begin{center}
\includegraphics[scale=0.55]{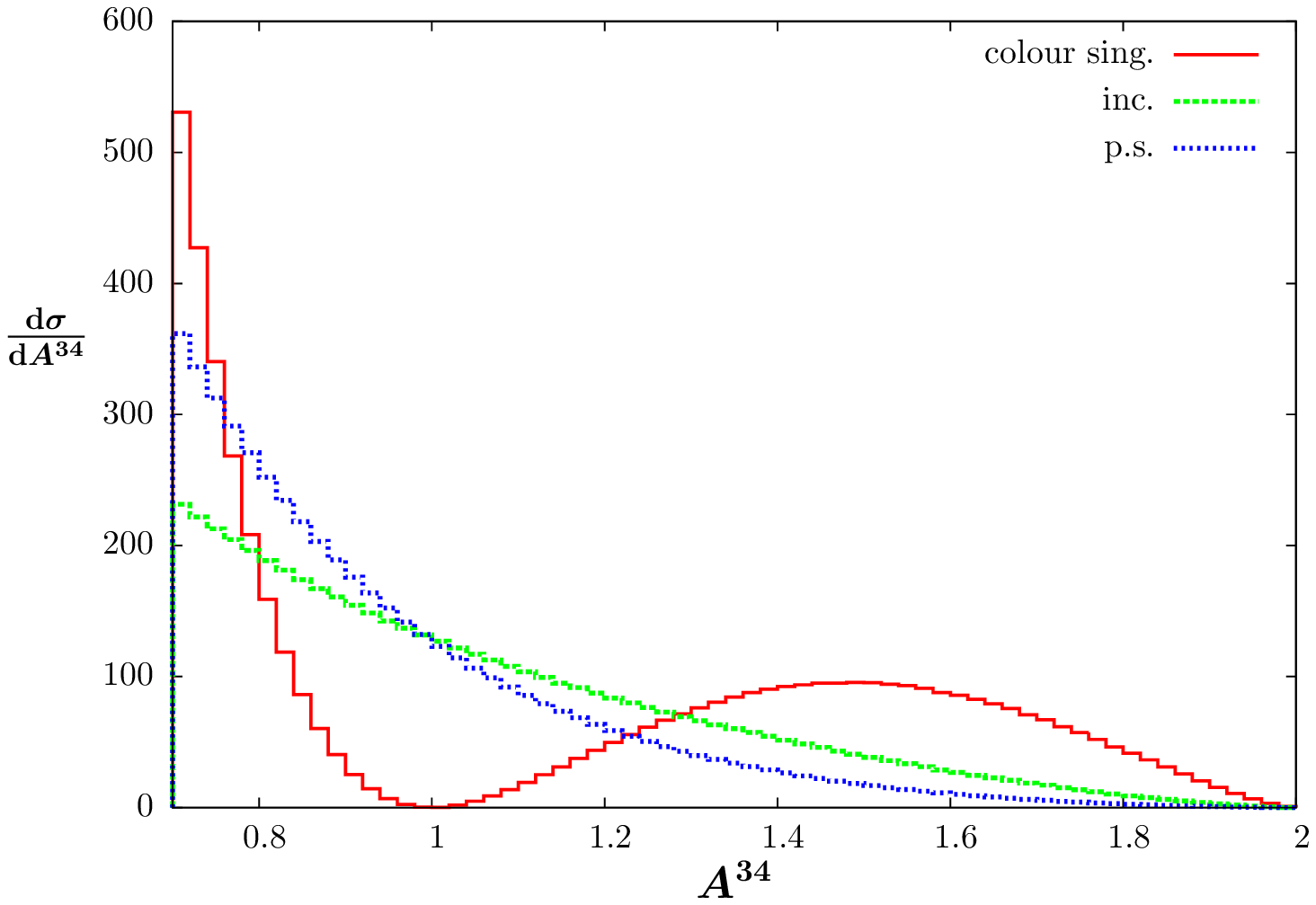}
\includegraphics[scale=0.55]{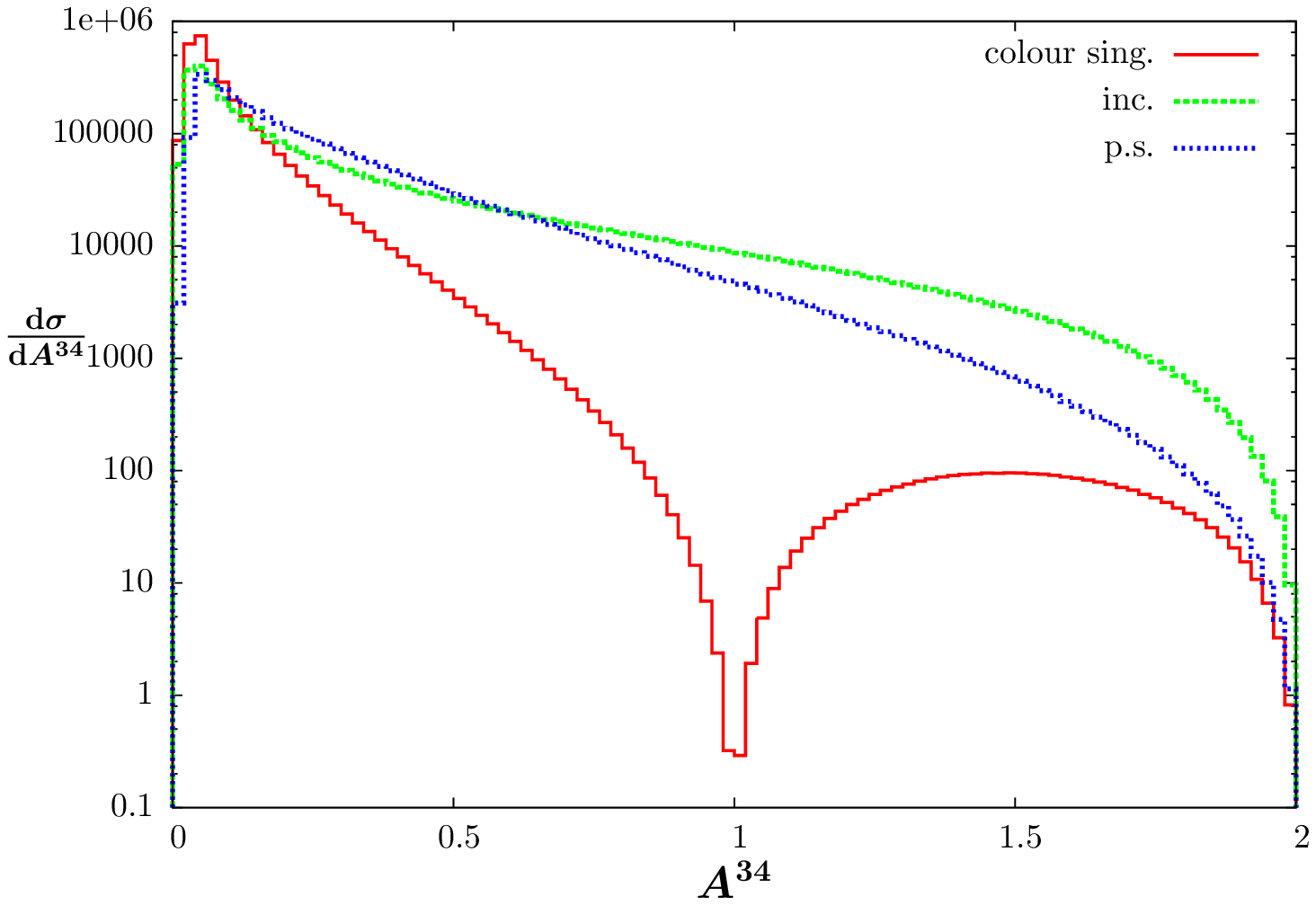}
\caption{Differential cross sections (in arbitrary units) with respect to the variable $A^{34}$, defined in the text, for 5--gluon scattering at tree--level, with the particle momenta restricted to lie in a plane. Plots are shown for the case of colour--singlet initial--state gluons, the inclusive colour averaged/summed case, and with the final--state particles distributed according to phase space. The integrated cross sections are normalized to each other in the region of each plot.}\label{aoplot}
\end{center}
\end{figure}

The radiation zeros are also exhibited in distributions with respect to the final--state parton momenta. Although these zeros, being relations between the particle rapidity differences, are generally not exhibited for a single value of a kinematic variable (e.g. a scattering angle), it is nonetheless possible to define suitable variables in which they can be observed. In all of the results which follow, the final--state partons are required to have transverse momentum $k_\perp>25$ GeV, and the $k_t$ algorithm with jet radius $R=0.6$ is applied to select three--jet events. All results are presented at parton--level and are intended for illustration; a complete phenomenological treatment would, for example, have to account for the non--trivial effect of parton shower and hadronization on the observables considered below.

Considering the 5--gluon colour--singlet amplitude, it is useful to define
\begin{equation}\label{A_0}
 A^{ij} \equiv \frac{{\rm sinh}^2\left(\frac{\Delta_{ij}}{2}\right)}{{\rm cosh}^2\left(\frac{\Delta_{jk}}{2}\right)+{\rm cosh}^2\left(\frac{\Delta_{ik}}{2}\right)}\;,
\end{equation}
where $(i,j,k)$ is a permutation of the gluon labels $(3,4,5)$. Thus, the zero condition (\ref{master1}) is satisfied when $A^{34}=1$, when as before the gluons 3 and 4 are by definition aligned. In Fig.~\ref{aoplot} we show the differential cross section with respect to $A^{34}$ for this process, subject to the condition that the gluons are in an exactly planar configuration. The distributions for the colour summed/averaged cross section, which contributes inclusively, as well as that due to phase space (i.e. with a uniform matrix element) are shown for comparison. The distinct behaviour of the colour--singlet cross section, and in particular the clear zero at $A^{34}=1$, is evident.

 \begin{figure}
\begin{center}
\includegraphics[scale=0.55]{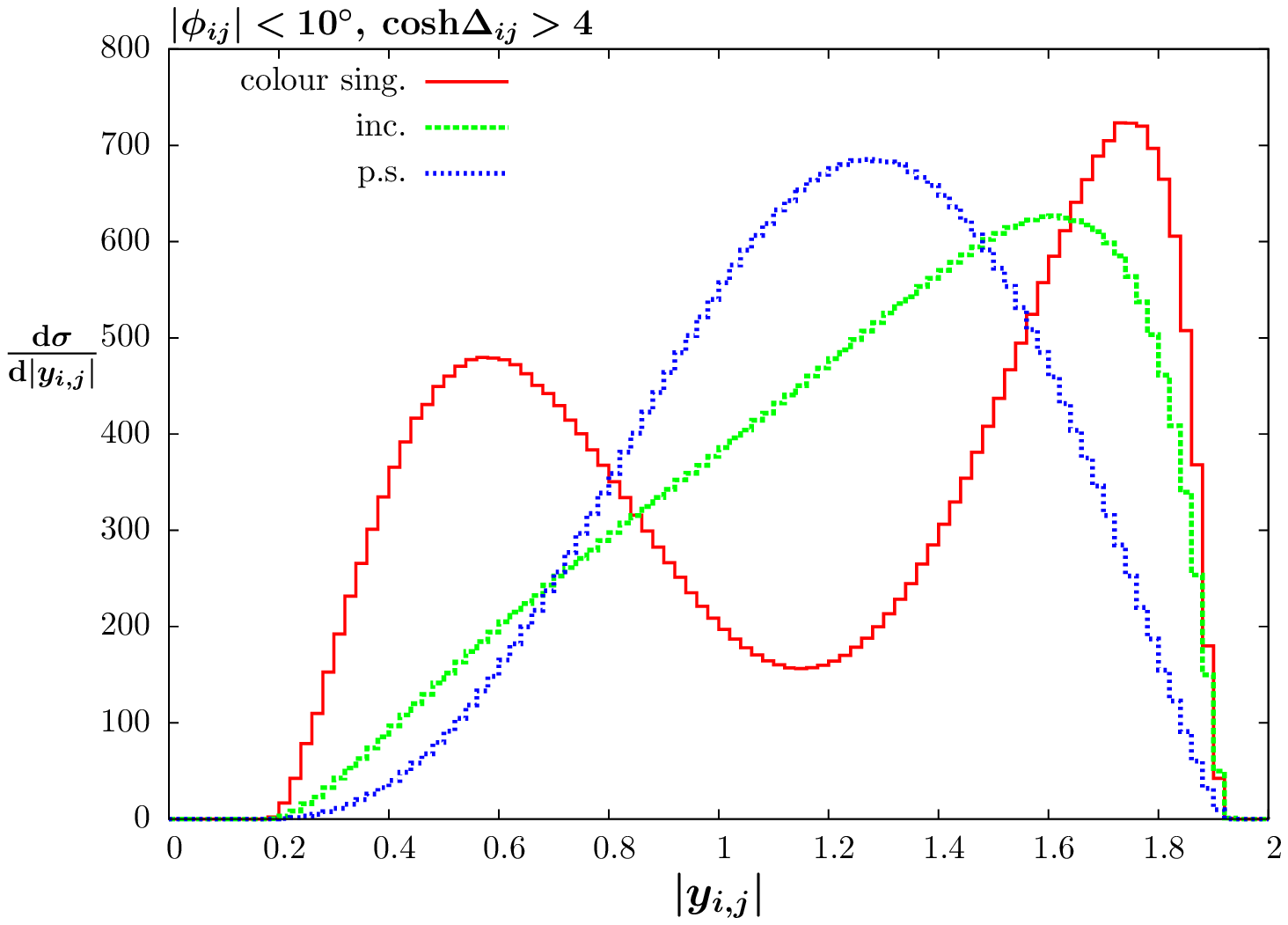}
\includegraphics[scale=0.55]{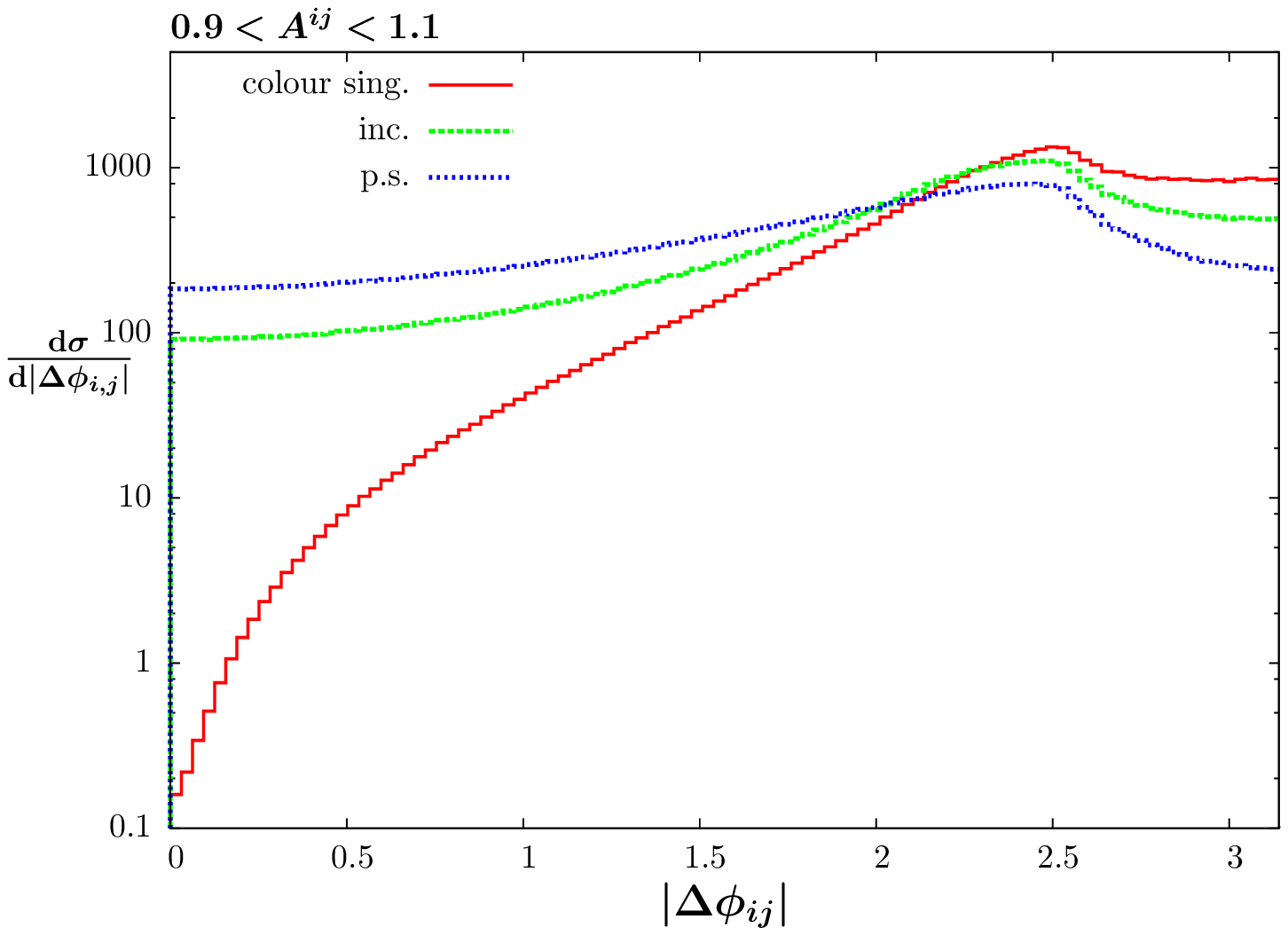}
\caption{Differential cross sections (in arbitrary units) for 5--gluon scattering at tree--level, with respect to: (left) the absolute value of the gluon rapidity $|y_{i,j}|$, with the cut $\Delta \phi_{ij} < 10^\circ$ and ${\rm cosh} \Delta_{ij}>4$ imposed; (right) The azimuthal angular separation $\Delta \phi_{ij}$, for gluon pairings passing the cut $0.9<A^{ij}<1.1$, where $A^{ij}$ is defined in (\ref{A_0}). Plots are shown for colour--singlet initial--state gluons, the inclusive colour averaged/summed case, and with the final--state particles distributed according to phase space. The integrated cross sections are normalized to each other in the region of each plot.}\label{y3zero}
\end{center}
\end{figure}

More realistically, events which are approximately in a planar configuration can be selected by imposing suitable cuts. In Fig.~\ref{y3zero} (left), the distribution with respect to the absolute value of the gluon rapidities $|y_{i,j}|$, subject to the requirement that $|\Delta \phi_{ij}| <10^\circ$ is shown, i.e. events are selected where one of the gluon pairings satisfies this constraint, and both gluon rapidities are then binned. Upon inspection it can be shown that (\ref{master1}) only has a solution for ${\rm cosh} \Delta_{34}>7$; we impose an additional, somewhat lower, cut of ${\rm cosh} \Delta_{ij}>4$ here to further isolate the kinematic region where a zero can occur without masking the dip structure by a more stringent, higher, cut. After this, although for the reasons discussed above a zero does not occur, a clear radiation dip is present in the resulting distribution. Comparing to the phase--space only and inclusive distributions, we can see that this is indeed driven by the zero condition, rather than being, say, an artefact of the cut choices. In Fig.~\ref{y3zero} (right) the distribution with respect to the angular separation $|\Delta\phi_{ij}|$ for gluon pairings passing the cut $0.9<A^{ij}<1.1$ is shown. A pronounced suppression for lower values of  $|\Delta\phi_{ij}|$, driven by the zero conditions (\ref{phireq}, \ref{master1}), is evident. Again, the behaviour of the inclusive and phase--space only distributions is completely different, with no such tendency to strongly disfavour lower $|\Delta\phi_{ij}|$ values.

Finally, it should be emphasised that the 5--gluon colour--singlet amplitude is the relevant object in the case of central exclusive trijet production, as discussed in the introduction, and as such these zeros represent physical observables in this process. However, for other specific colour choices, such as those taken for demonstration in Fig.~\ref{y34scan} (right), the configuration has no observable relevance; in the inclusive cross section, it is the squared amplitude, summed/average over all colours, which contributes. It is therefore worth considering briefly whether these planar zeros manifest themselves in this inclusive cross section. As the form of the zero curves (and indeed whether any solution to (\ref{master1g}) exists at all) shown in Fig.~\ref{y34scan} depends strongly on the colour configuration, it is immediately apparent that no exact zero will remain in the inclusive cross section; however, it is at least in principle possible that a radiation dip structure may remain.
 The simple form of (\ref{master1g}) allows a relatively straightforward expressions to be written down for these when the partons are in a planar configuration: these are given in Appendix~\ref{inc} for the representative 5--gluon and $q\overline{q}\to ggg$ processes. Although the form of these cross sections do not completely rule out such a dip structure, no clear evidence of this is found. 

\subsection{Non--planar zeros}\label{secnonplanar}

So far, we have only considered the planar case, however, as noted in Section~\ref{secgenc}, there is in principle another possible way that a zero may occur, without such a requirement on the particle orientation: in particular, if the two conditions (\ref{otherzero}) are individually satisfied, neither of which will in general require  a planar configuration. 

To clarify this further we can consider the 5--gluon amplitude for the case of colour--singlet inital--state gluons, as in Section~\ref{firstlook}. Using (\ref{prodex}), but without  making any assumptions about the azimuthal orientation of the particles, it can be shown that the general conditions for the existence of a zero are given by
\begin{align}\label{gcond1}
\cos \Delta \phi_{45}\,{\rm cosh}\Delta_{45}+\cos \Delta \phi_{34}\,{\rm cosh}\Delta_{34}+\cos \Delta \phi_{35}\,{\rm cosh}\Delta_{35}&=3\;,\\ \label{gcond2}
\sin \Delta\phi_{45}\, {\rm sinh}\Delta_{45}+\sin \Delta\phi_{34}\, {\rm sinh}\Delta_{34}+\sin \Delta\phi_{35}\, {\rm sinh}\Delta_{35}&=0\;,
\end{align}
while expressions for general colour are given by (\ref{gcond1}) and (\ref{gcond2}) in Appendix~\ref{apnonplanar}. We can see that (\ref{gcond2}) is automatically satisfied if the particles are in a planar configuration, and in this case (\ref{gcond1}) reduces to (\ref{master1}) for the corresponding gluon alignment, as it must. However it appears to be quite possible for these constraints to both be satisfied without such a requirement. 
 \begin{figure}
 \begin{center}
 \includegraphics[scale=0.55]{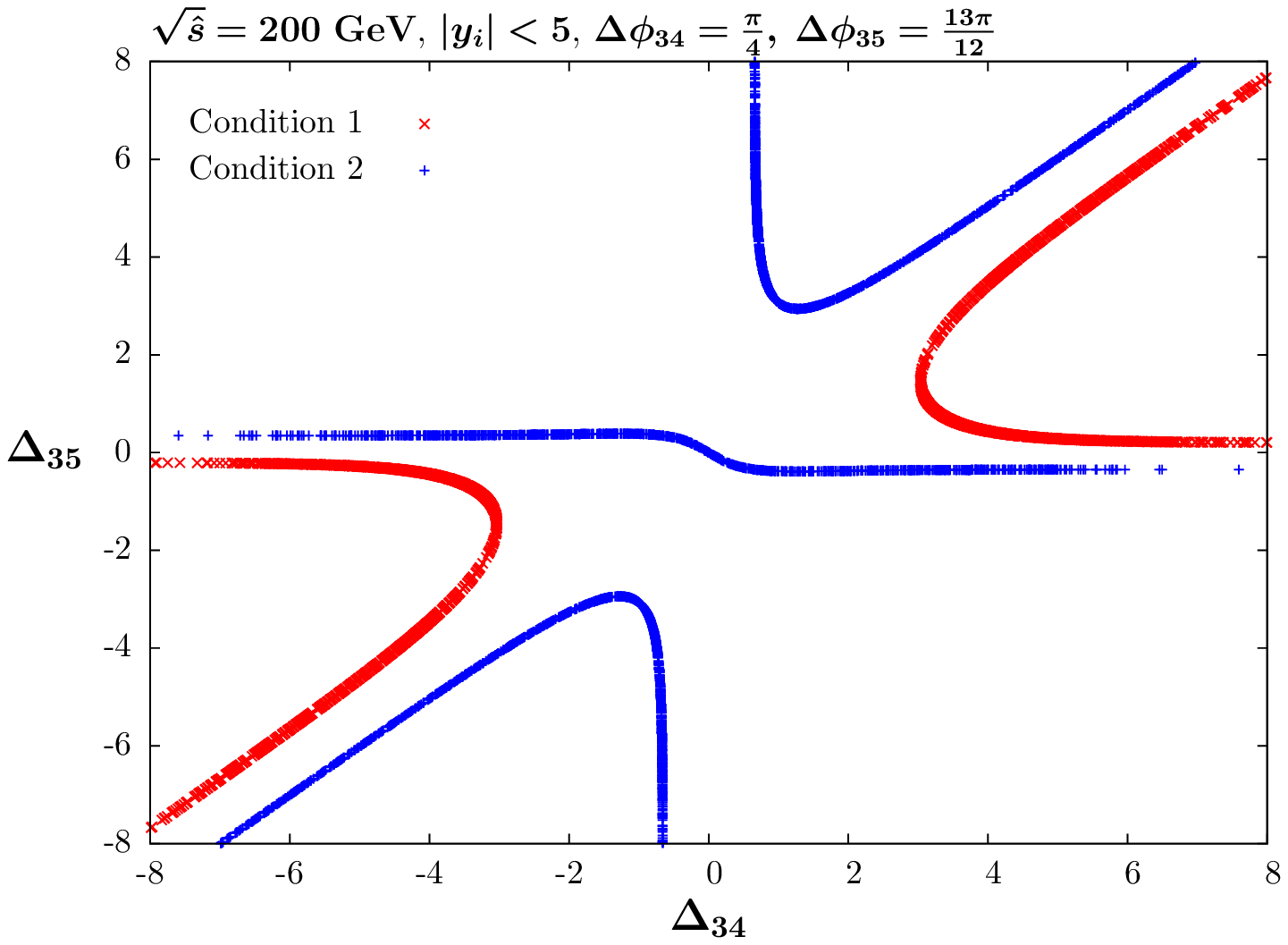}
  \includegraphics[scale=0.55]{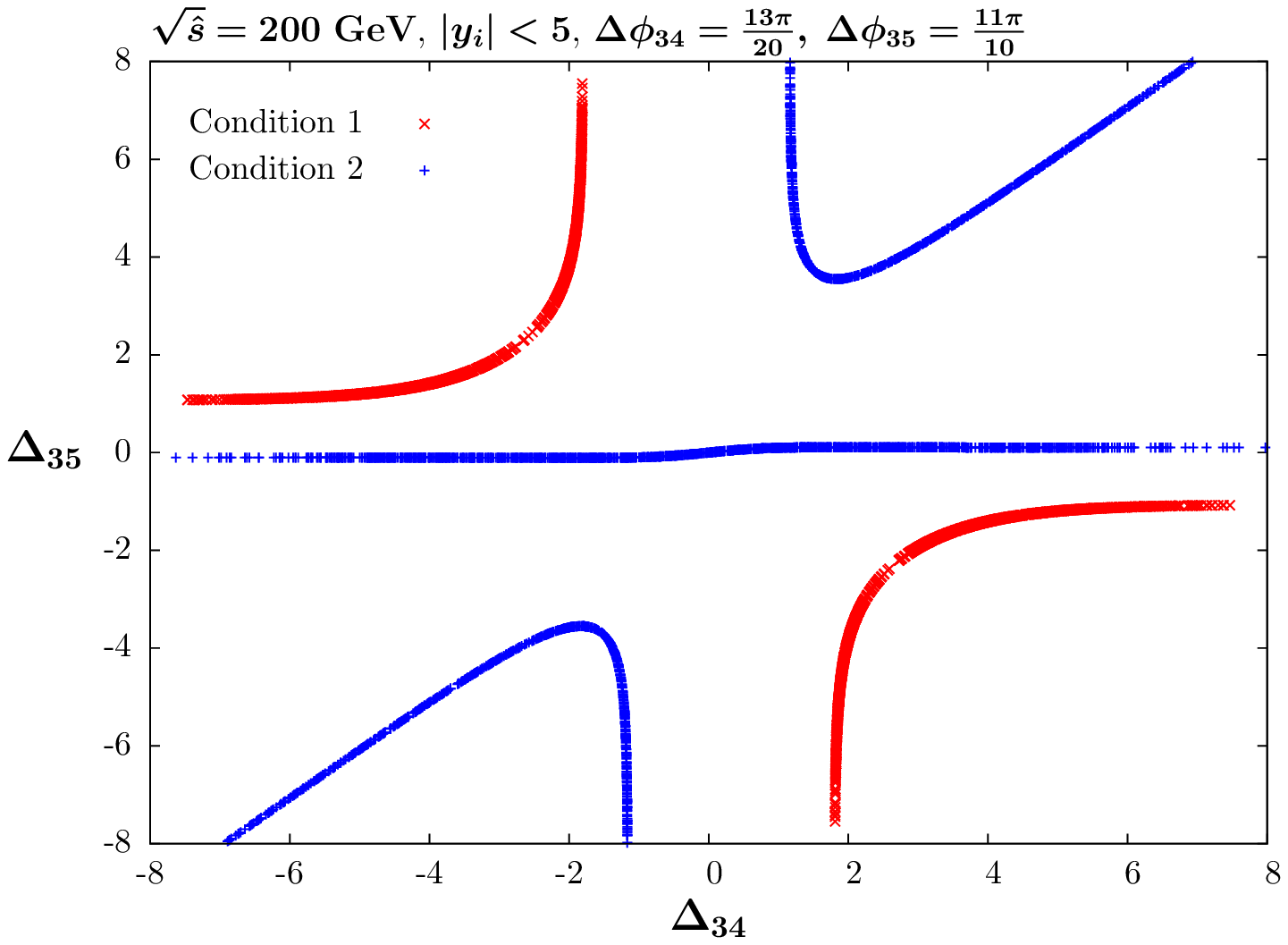}
 \caption{Scan of solutions to (\ref{gcond1}) and (\ref{gcond2}), labelled `Condition 1' and `Condition 2', respectively, which lie in the physical region for the colour--singlet 5--gluon amplitudes, as a function of the rapidity differences $\Delta_{34}$ and $\Delta_{45}$. The central system mass is $\sqrt{\hat{s}}=200$ GeV, and the final--state particles are required to have rapidity $|y_i|<5$. The plots correspond to two different choices of angular separations, $\Delta \phi_{ij}$, with the values indicated in the figures.}\label{y34scangen}
 \end{center}
 \end{figure}
To investigate whether this is the case, we can consider fixed values for the angular separations $\Delta\phi_{34}$, $\Delta\phi_{35}$. The labelling here is arbitrary and has no physical interpretation, as there is no longer any requirement that gluons 3 and 4 are aligned. It is only specified here for clarity; the physically relevant feature is the correspondence between the angular separation $\Delta \phi_{ij}$ and the rapidity difference $\Delta_{ij}$ in Fig.~\ref{y34scangen}. We can then use momentum conservation to define one particle rapidity, $y_i$, via
\begin{equation}\label{momcons}
\sin \Delta\phi_{35}\,{\rm sinh}\, y_4-\sin\Delta\phi_{34}\,{\rm sinh}\, y_5=\sin\Delta\phi_{45}\,{\rm sinh} \,y_3\;,
\end{equation}
with the remaining three constraints defining the parton transverse momenta. We then show in Fig.~\ref{y34scangen}, the regions of ($\Delta_{34}$, $\Delta_{35}$) space  where (\ref{gcond1}) and (\ref{gcond2}) are satisfied, for two representative choices of angular separation $\Delta \phi_{34}$, $\Delta \phi_{35}$. In both cases, we can see that these can be individually satisfied for a range of rapidity differences, $\Delta_{ij}$; however, the corresponding curves where the zero conditions are satisfied are completely non--overlapping. 
A similar effect is found for other choices of $\Delta\phi_{ij}$, suggesting that both (\ref{gcond1}) and (\ref{gcond2}) cannot be simultaneously satisfied. This is confirmed by a precise numerical scan of the corresponding scattering amplitude over the full phase space, and the same result is found for the $gg \to q\overline{q}g$ amplitude with colour--singlet initial--state gluons.

\begin{figure}
 \begin{center}
 \includegraphics[scale=0.55]{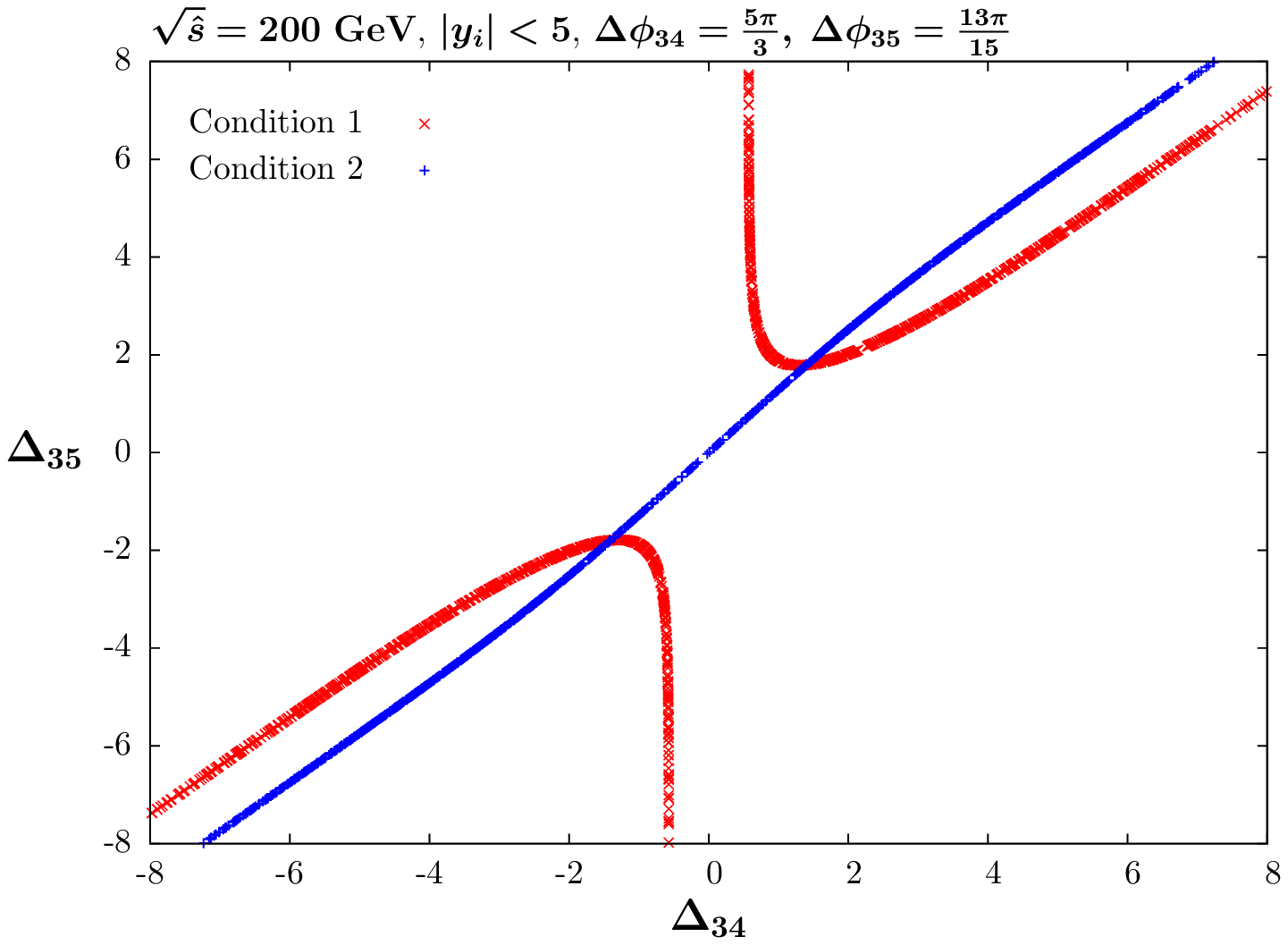}
  \includegraphics[scale=0.55]{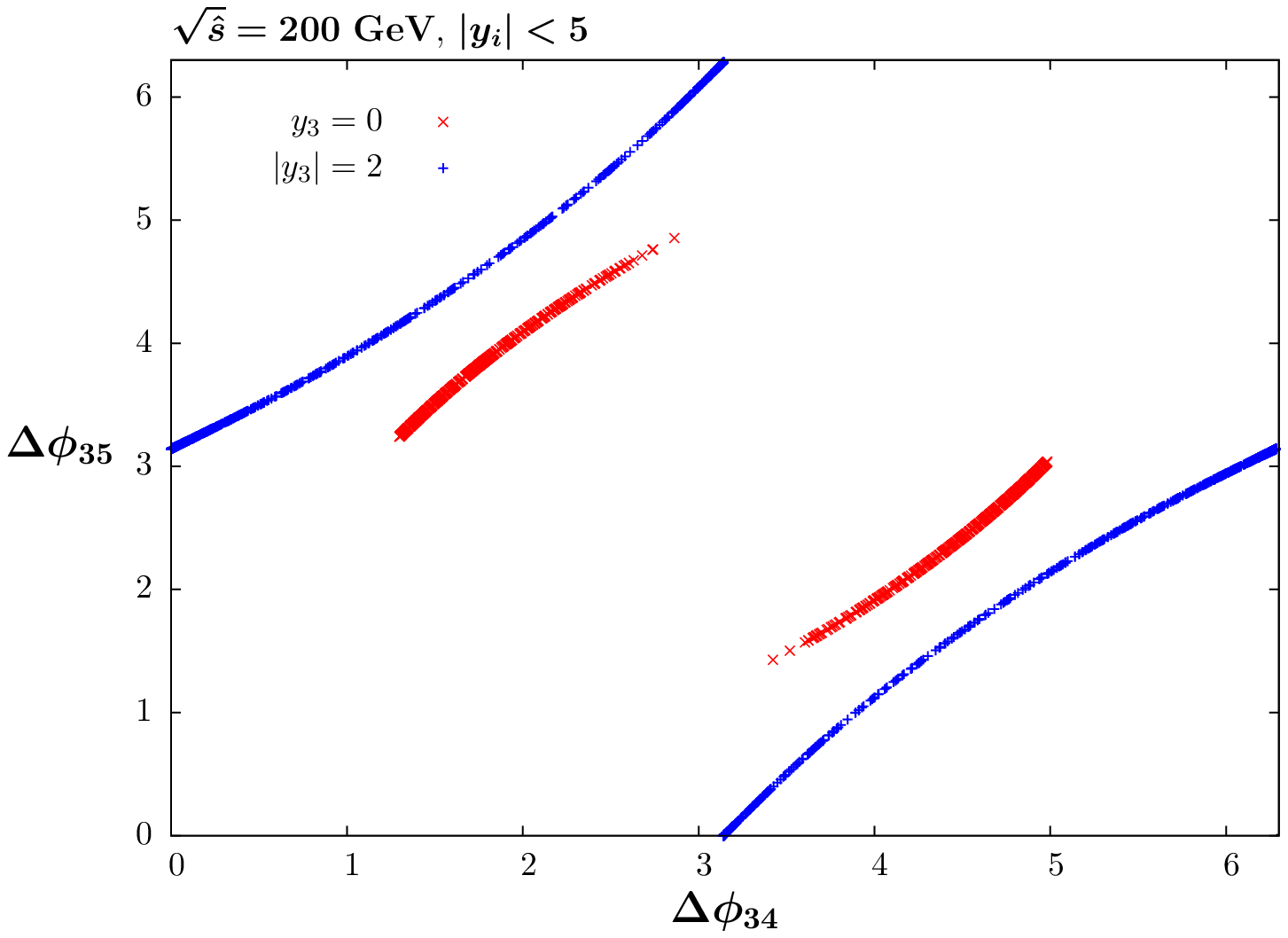}
 \caption{(Left) Scan of solutions to (\ref{npc1}) and (\ref{npc2}), labelled `Condition 1' and `Condition 2', respectively, which lie in the physical region for the 5--gluon amplitudes, as a function of the rapidity differences $\Delta_{34}$ and $\Delta_{45}$. (Right) Scan of solutions to both (\ref{npc1}) and (\ref{npc2}), as a function of the azimuthal angular separations $\Delta\phi_{34}$ and $\Delta\phi_{45}$, for fixed values of rapidity $y_3$. Plots are shown for the specific choice of colour indices $(c_1,c_2,c_3,c_4,c_5)=(7,7,3,4,5)$. The central system mass is $\sqrt{\hat{s}}=200$ GeV, and the final--state particles are required to have rapidity $|y_i|<5$.}\label{y34scangen1}
 \end{center}
 \end{figure}

On the other hand, there are a wide range of other colour configurations for which non--planar zeros may still be possible. After a detailed numerical scan over all colours for solutions to (\ref{npc1}) and (\ref{npc2}), it is found that indeed such zeros do occur in a range of cases, albeit a smaller number than for the planar zeros. A representative example is given in Fig.~\ref{y34scangen1}, which corresponds to the 5--gluon amplitude for a specific choice of colour indices $c_i$. In Fig.~\ref{y34scangen1} (left) the equivalent plot to Fig.~\ref{y34scangen} is given, produced as described above: in contrast to the colour--singlet case we can see that the zero curves for the two conditions clearly intersect. At these intersection points both conditions are satisfied and non--planar zeros will exist. To demonstrate this further we show in Fig.~\ref{y34scangen1} (right) the scan of simultaneous solutions to both conditions as a function of $\Delta\phi_{ij}$. Accounting for momentum conservation and one of the zero conditions, we are left with two independent azimuthal differences, and one particle rapidity, $y_3$, say. The remaining condition therefore defines a volume in $\Delta\phi_{ij}$ space, or a curve for fixed $y_3$. Two such curves, for particular choices of $|y_3|$ (for the particular colour choice the solutions are symmetric in $y_i \to -y_i$) are shown in the figure, and the range of solutions for non--planar configurations is clear.

Finally, it can be checked numerically that a condition of the kind given by (\ref{type1}) does not hold: by inspection it is found that the ratios $s_{ij}/s_{ik}$ are generally not constant when these zeros occur. As these are self--evidently not of type--II, being non--planar, the nature of these zeros is therefore not completely clear. While these only occur for specific, unobservable, colour configurations of the scattering particles, and therefore have no direct phenomenological consequence, it would nonetheless be interesting to find a more general classification for these zeroes, if it exists, as well as to determine whether they occur in other (e.g. QED) processes. However, the emphasis of this paper is on the planar zeros discussed in the previous sections, and a full investigation of these issues is left to future work.

\section{Summary and outlook}\label{conclusions}

Radiation zeros are an interesting effect whereby the tree--level scattering amplitudes involving the radiation of one or more massless gauge bosons can completely vanish for particular configurations of the final--state particles. While the general conditions for a wide class of such zeros have been found long ago~\cite{Brown:1982xx,Brodsky:1982sh,Samuel:1983eg}, subsequently an additional type of zero was discovered in~\cite{Heyssler:1997fy}. In this case, the amplitude is found to vanish if and only if the particle 3--momenta lie in a plane, and satisfy an additional kinematic condition that depends on the scattering process. These `planar', or `type--II', radiation zeros have been observed in a small group of electroweak and QED scattering processes~\cite{Heyssler:1997fy,Stirling:1999sj,Rodriguez:2003tt}, but remain relatively unstudied, and the general principles for their occurrence have yet to be truly understood. 

In this paper we have demonstrated, for the first time, the existence of planar radiation zeros in QCD processes for general momenta.  We have considered all 5--parton tree--level amplitudes, that is $gg \to ggg$, $q\overline{q}  \to ggg$ and $q\overline{q}  \to q\overline{q} g$ and those related by  initial/final--state crossings, and shown that in all cases planar zeros exist for particular colour configurations. This fact is shown to follow particularly simply when the BCJ relations~\cite{Bern:2008qj}, which allow the $n$--parton QCD amplitudes to be written as a linear combination of only $(n-3)!$ independent partial amplitudes, are used. Making further use of the MHV formalism, we have then derived simple expressions, involving only the rapidity differences of the final--state partons, for the remaining kinematic condition for these zeros. The simplicity of these conditions, which express a delicate cancellation between a large number of contributing Feynman diagrams, relies on the simplification allowed by the MHV approach and the BCJ relations. It is worth mentioning in passing that many of the general analyses of radiation zeros were performed before the development of the MHV formalism; it is therefore possible that this approach has yet to be fully exploited, in  the case of both QED and QCD radiation zeros.


Numerical results demonstrating that these planar zeros do indeed occur in the physical region for a wide range of amplitudes have been presented, and in addition it has been shown that zeros can occur for non--planar configurations of the 5--parton QCD amplitudes, but which do not generally satisfy the type--I constraint (\ref{type1}). While this paper has concentrated on the known planar case, the nature of these non--planar zeros, and in particular whether they can be understood or classified in a more general way, requires further study.

Other avenues of  future investigation include the possibility of a generalisation to higher points. Indeed, in a recent work~\cite{Harland-Lang:2014efa}  a radiation zero is found to exist in the $gg \to J/\psi J/\psi$ subprocess, i.e. in the 6--parton $gg\to q\overline{q} q\overline{q}$ amplitudes where the collinear $q\overline{q}$ pairs form the parent $J/\psi$ meson. While these zeros appear to be helicity--dependent, and the contributing diagrams are also determined by the odd $C$--parity of the $J/\psi$, it is nonetheless worth emphasising that such a collinear configuration  automatically corresponds to a planar orientation in the $6$--parton amplitude. In addition, the amplitudes in which these zeros occur are for massive (charm) quarks; it would be interesting to see the effect of including a quark mass in the processes considered  in this work.  An equivalent zero is also found in~\cite{HarlandLang:2011qd} for the $gg \to \pi\pi$ process calculated within the `hard exclusive' formalism~\cite{Lepage:1980fj}; while here there is no explicit helicity dependence, again only a subset of Feynman diagrams contribute due to the quantum numbers of the produced mesons. More generally, there is no reason to believe that  planar (and non--planar) zeros do not occur in the $n=6,7...$ parton amplitudes: an investigation of the $n\geq 6$ parton amplitudes, making use of the simplification allowed by the BCJ relations, would clarify this. 

Finally, while the form of these zeros generally depends on the unobservable colour indices of the scattered particles,  a planar zero has been shown to exist when the incoming gluons are in a colour--singlet state, both for the 5--gluon and, in certain circumstances, $gg \to q\overline{q} g$ amplitudes. These are precisely the amplitudes which contribute to central exclusive  three--jet  production, and therefore the effect of such zeros may be observable in this channel: a detailed phenomenological study is planned in~\cite{HKRfut}.

\section*{Acknowledgements}

I thank Valery Khoze, Misha Ryskin and James Stirling for useful discussions, and the Science and Technology Facilities Council (STFC) for support via the grant award ST/L000377/1. I also acknowledge the support of the IPPP, Durham, where the initial stages of this study were performed.

\appendix

\section{Spinor identities}\label{app:spinor}

Here, some useful identities and relations satisfied by the spinor contractions $\langle k_i\,k_j\rangle$ are given. For the Dirac representation of the $\gamma$ matrices, these are given by~\cite{Mangano:1990by}
\begin{equation}\label{cont}
 \langle k_i\,k_j\rangle =\sqrt{k_i^-k_j^+}e^{i\phi_i}-\sqrt{k_i^+k_j^-}e^{i\phi_j}\;,
\end{equation}
where $k^\pm = k^0 \pm k^3$ and $\phi_j$ is the azimuthal angle in the $x$--$y$ plane, i.e.
\begin{equation}
 e^{\phi_j}=\frac{k_{j}^x+ik_{j}^y}{\sqrt{k_j^+k_j^-}}\;.
\end{equation}
Writing (\ref{cont}) in terms of explicit component, we find
\begin{equation}\label{prodex}
 \langle k_i\,k_j\rangle = 2\,(k_\perp^i k_\perp^j)^{1/2} e^{i\frac{\phi_i+\phi_j}{2}}\left(i\sin \frac{\Delta \phi_{ij}}{2}\,{\rm cosh}\frac{\Delta_{ij}}{2}-\cos \frac{\Delta \phi_{ij}}{2}\,{\rm sinh}\frac{\Delta_{ij}}{2}\right)\;,
\end{equation}
when both particles $i$ and $j$ are in the final state. Here $\Delta_{ij}=y_i-y_j$ and $\Delta \phi_{ij}=\phi_i-\phi_j$ are the difference in rapidity and azimuthal angle between the two particles, respectively, and $k_\perp$ is the particle transverse momentum. Finally, the spinor products satisfy the Schouten identity
\begin{equation}\label{sc}
\langle i\,j\rangle \langle k\,l\rangle =\langle i\,l\rangle \langle k\,j\rangle +\langle i\,k\rangle \langle j\,l\rangle\;,
\end{equation}
while momentum conservation gives
\begin{equation}\label{cons}
\sum_{\substack{i=1\\  i\neq j,k}}^n s_{ij}\frac{\langle i\, k \rangle}{\langle i\, j \rangle}=0\;,
\end{equation}
where $s_{ij}=(k_i+k_j)^2$.

\section{General zero conditions}\label{apnonplanar}

Here we give the general conditions for the existence of a radiation zero in the 5--gluon and $q\overline{q}\to ggg$ amplitudes, as discussed in Section~\ref{secnonplanar}. These are
\begin{align}\nonumber
&\cos \Delta \phi_{45} \left(\tilde{C}_{45}\,{\rm cosh}\Delta_{45}+C_{45}\, {\rm sinh}\Delta_{45}\right)-\cos\Delta\phi_{35}\left(\tilde{C}_{35} \,{\rm cosh}\Delta_{35}+C_{35}\,{\rm sinh}\Delta_{35}\right)\\ \label{npc1}
&+\cos\Delta\phi_{34}\left(\tilde{C}_{34}\,{\rm cosh}\Delta_{34}+C_{34}\,{\rm sinh}\Delta_{34}\right)+\left(\tilde{C}_{35}-\tilde{C}_{45}-\tilde{C}_{34}\right)=0\;,
\end{align}
and
\begin{align}\nonumber
&\sin \Delta \phi_{45} \left(\tilde{C}_{45}\,{\rm sinh}\Delta_{45}+C_{45}\, {\rm cosh}\Delta_{45}\right)-\sin\Delta\phi_{35}\left(\tilde{C}_{35} \,{\rm sinh}\Delta_{35}+C_{35}\,{\rm cosh}\Delta_{35}\right)\\ \label{npc2}
&+\sin\Delta\phi_{34}\left(\tilde{C}_{34}\,{\rm sinh}\Delta_{34}+C_{34}\,{\rm cosh}\Delta_{34}\right)=0\;,
\end{align}
where the colour factors are defined as in (\ref{c45}) and (\ref{c45q}) for the gluon and quark amplitudes, respectively. 

\section{Colour--summed squared amplitudes}\label{inc}

Here we give the explicit form for the non--trivial rapidity dependence of the squared matrix elements, suitably average/summed over colour indices, in the case of the $gg \to ggg$ and $q\overline{q} \to ggg$ scattering processes, for planar parton configurations. These can be derived in a relatively straightforward fashion from (\ref{master1g}), in particular once it is observed that any terms proportional to $\tilde{C}_{ij}C_{ij}^\star$ vanish upon summing over all colours. We have
\begin{align}\nonumber
|\overline{\mathcal{M}}|^2 &\propto n_1\left({\rm sinh}^2 \Delta_{45}+{\rm sinh}^2 \Delta_{35} +{\rm sinh}^2 \Delta_{34} \right)
+ n_2\left({\rm cosh}^4 \frac{\Delta_{45}}{2}+{\rm cosh}^4 \frac{\Delta_{35}}{2}+{\rm sinh}^4 \frac{\Delta_{34}}{2}\right) \;,\\ \nonumber
&+ n_3\left({\rm sinh} \Delta_{35}\,{\rm sinh} \Delta_{34}-{\rm sinh} \Delta_{45}\,{\rm sinh} \Delta_{34}-{\rm sinh} \Delta_{45}\,{\rm sinh} \Delta_{35}\right)\;,\\
&+ n_4\left({\rm cosh}^2 \frac{\Delta_{45}}{2}\,{\rm sinh}^2 \frac{\Delta_{34}}{2}+{\rm cosh}^2 \frac{\Delta_{35}}{2}\,{\rm sinh}^2 \frac{\Delta_{34}}{2}-{\rm cosh}^2 \frac{\Delta_{45}}{2}\,{\rm cosh}^2 \frac{\Delta_{35}}{2}\right)\;,
\end{align}
where overall kinematic and colour factors are omitted, and in both cases gluons 3 and 4 are by definition aligned. In the gluon case we have
\begin{equation}
n_1^g=2,\quad n_2^g=8,\quad n_3^g=3,\quad n_4^g=4\;,
\end{equation}
and in the $q\overline{q} \to ggg$ case we have 
\begin{equation}
n_1^q=\frac{1}{4}\left(N_c^2-1+\frac{2}{N_c^2}\right), \quad n_2^q=N_c^2-3,\quad n_3^q=-\frac{N_c^2-2}{2 N_c^2},\quad n_4^q = -2\;.
\end{equation}

\bibliography{ggbib}{}
\bibliographystyle{h-physrev}

\end{document}